\documentclass[10pt,twocolumn]{article}

\usepackage{graphics,cite,subfigure}
\usepackage[left=1.5cm,right=1.5cm,top=1.5cm,bottom=1.5cm]{geometry}
\begin{document}

\title{Reconstructing signal from fiber-optic measuring\\ system with 
non-linear perceptron}

\author{A. V. Panov\protect\thanks{{\it\ E-mail:} panov@iacp.dvo.ru} \\[1ex]
        \textit{Institute for Automation and Control Processes, 
	Russian Academy of Sciences}\\ \textit{Far-Eastern Branch, 
	5, Radio st., Vladivostok, 690041, Russia}
\\[12pt]}
\date{}

\maketitle

\begin{abstract} 

A computer model of the feed-forward neural network with the hidden layer  is
developed to reconstruct physical field investigated by the fiber-optic
measuring system. The Gaussian distributions of some physical quantity are
selected as learning patterns. Neural network is learned by error
back-propagation using the conjugate gradient and coordinate descent
minimization of deviation. Learned neural network reconstructs the 
two-dimensional scalar physical field with distribution having one or two 
Gaussian peaks.\par

\end{abstract}

\section{Introduction}

The fiber-optic distributed measuring systems were being actively developed
during the past few years \cite{quant97,opmem97,fiber98}.  Particularly
these systems can be used for acoustic field investigating, for monitoring of
stresses in industry, in geophysical researches, etc. The reconstruction of
information obtained by measuring system needs solving of tomographic problem.
But frequently such solution is complex, needs large computer consumption and
cannot be performed in real time. Therefore previously learned neural network
can be utilized for solving of tomographic problem \cite{opmem97,cich}.

An optical perceptron, which was implemented using a collection of amplitude
holograms recorded on a disk-shaped holographic carrier and processing the
output data from a fiber-optic  measuring system, is reported in Ref.
\cite{zhtf99}. Another computer linear neural network was developed by authors
of Refs. \cite{psr99,lb95}.  We present the similar computer non-linear
feed-forward neural network in this paper.

\section{Description of the neural network}

\begin{figure}
\centerline{\includegraphics{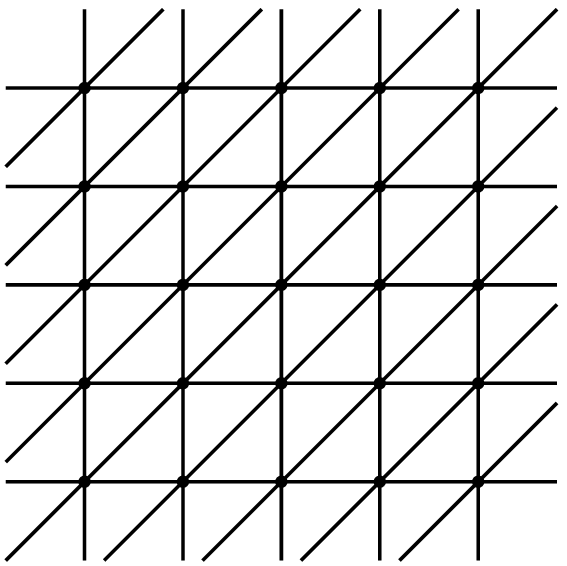}}

\caption{\label{square} The $5\times 5$ fiber-optic distributed measuring system.}
\end{figure}

Consider two-dimensional fiber-optic distributed measuring system being set of
measuring lines. The fiber-optic phase sensors form square $n\times n$ lattice
also having $2n-1$ diagonal lines (Fig. \ref{square}).  The
phase shift $\theta$ in this linear sensor is proportional to physical action
value integrated over the line while the intensity of the transmitted light
$\propto \cos^2\theta$.  The phase shift in fiber-optic line have to be along
one cosine half-period. Signal from such measuring system is simulated on input
of neural network which is intended to reconstruct the value of physical action
at intersection points.

For solving of tomographic problem we choose perceptron with the non-linear
hidden layer since this network has universal approximation capability
\cite{funa,hornik}.

The neurons of the first layer serve as network inputs and feed data from the
measuring system to the next layer.
The second (hidden) layer proceeds following transformation:
\begin{equation}
s_j=f\Bigl( \sum_k \overline w_{jk}\sigma_k\Bigr),
\end{equation}
where $\sigma_k$ are states of neuron inputs being signals from measuring 
sensors, $s_j$ are states of outputs and 
$\overline w_{jk}$ are synapses. We used $\tanh$ as activation function $f$.
The hidden layer of neurons has $4n-1$ inputs and $n\times n$ outputs.

The output layer of the neurons takes the linear transformation:
\begin{equation}
S_j=\sum_j w_{ij}s_j,
\end{equation}
where $S_j$ are states of third layer neuron outputs, $w_{ij}$ are synapses of
third layer.

We selected Gaussian distributions of the some 
physical quantity with one or two peaks as learning
patterns. 
The following expression was used as objective function:
\begin{equation}
D = \frac 1 2 \sum _{\mu, i}\left(S_i^\mu - \zeta _i^\mu \right)^2,
\end{equation}
where $\mu$ is superscript indicating number of learning pattern, 
$\zeta_i$ are output states of the neural network for some learning pattern.
For the training pattern $\sigma_k$ are formed as squared cosines of the sum 
of $\zeta_i$ along the
corresponding measuring line.
We used error back-propagation for the network training, so we had to minimize
$D$ with respect to $\overline w_{jk}$, $w_{ij}$.

We combined conjugate gradient and coordinate descent minimization methods for
neural network learning. When conjugate gradient method failed we switched
to the coordinate descent and after some iterations tried to start
conjugate gradient minimization again. Also we applied the ``thermal jumps''
technique. The minimization procedure finished after certain iteration
count or on given accuracy reaching.

\section{Results}

\begin{table}
\caption{\label{example-delta} The initial (a) and reconstructed by the neural
network (b) 
$\delta$-like distribution of physical quantity with one peak. Each cell corresponds to intersection point within
measuring system (see Fig. \ref{square}) and represents certain $\zeta_i$.}

\begin{center}

\subtable[]{
\begin{tabular}{rrrrr}
\hline
  0& 0& 0& 0& 0\\                                   
  0& 0& 0& 0& 0\\                                   
  0& 0.00005& 0.00327& 0.00005& 0\\                                   
  0& 0.00327& 0.20712& 0.00327& 0\\                                   
  0& 0.00005& 0.00327& 0.00005& 0\\                                   
\multicolumn{1}{c}{\rule{3.5em}{0ex}} & \multicolumn{1}{c}{\rule{3.5em}{0ex}} & 
\multicolumn{1}{c}{\rule{3.5em}{0ex}} & \multicolumn{1}{c}{\rule{3.5em}{0ex}} & 
\multicolumn{1}{c}{\rule{3.5em}{0ex}} \\[-\arraystretch pc]
\hline
\end{tabular}
}

\subtable[]{
\begin{tabular}{rrrrr}
\hline
  0.00001& 0.00063& 0.00027&-0.00003& 0.00003\\                                   
  0.00119& 0.00080&-0.00022& 0.00044&-0.00052\\                                   
 -0.00021& 0.00044& 0.00397&-0.00242& 0.00077\\                                   
  0.00071& 0.00117& 0.21311& 0.00188& 0.00113\\                                   
 -0.00097&-0.00081& 0.00274&-0.00012& 0.00063\\                                   
\multicolumn{1}{c}{\rule{3.5em}{0ex}} & \multicolumn{1}{c}{\rule{3.5em}{0ex}} & 
\multicolumn{1}{c}{\rule{3.5em}{0ex}} & \multicolumn{1}{c}{\rule{3.5em}{0ex}} & 
\multicolumn{1}{c}{\rule{3.5em}{0ex}} \\[-\arraystretch pc]
\hline
\end{tabular}
} 

\end{center}
\end{table}

\begin{table}
\caption{\label{example-smooth} The initial (a) and reconstructed by the neural
network (b) 
smooth distribution of physical quantity with one peak.}

\begin{center}
\subtable[]{
\begin{tabular}{rrrrr}
\hline
  0.00186&  0.00283&  0.00186&  0.00052&  0.00006\\
  0.01534&  0.02341&  0.01534&  0.00432&  0.00052\\
  0.05448&  0.08312&  0.05448&  0.01534&  0.00186\\
  0.08312&  0.12681&  0.08312&  0.02341&  0.00283\\
  0.05448&  0.08312&  0.05448&  0.01534&  0.00186\\
\multicolumn{1}{c}{\rule{3.5em}{0ex}} & \multicolumn{1}{c}{\rule{3.5em}{0ex}} & 
\multicolumn{1}{c}{\rule{3.5em}{0ex}} & \multicolumn{1}{c}{\rule{3.5em}{0ex}} & 
\multicolumn{1}{c}{\rule{3.5em}{0ex}} \\[-\arraystretch pc]
\hline
\end{tabular}
}

\subtable[]{
\begin{tabular}{rrrrr}
\hline
  0.00199&  0.00240&  0.00034&  0.00064& -0.00028\\
  0.01396&  0.02222&  0.01447&  0.00485&  0.00030\\
  0.05322&  0.07995&  0.05383&  0.01536&  0.00183\\
  0.08341&  0.12471&  0.08357&  0.02404&  0.00435\\
  0.05421&  0.08396&  0.05473&  0.01380&  0.00243\\
\multicolumn{1}{c}{\rule{3.5em}{0ex}} & \multicolumn{1}{c}{\rule{3.5em}{0ex}} & 
\multicolumn{1}{c}{\rule{3.5em}{0ex}} & \multicolumn{1}{c}{\rule{3.5em}{0ex}} & 
\multicolumn{1}{c}{\rule{3.5em}{0ex}} \\[-\arraystretch pc]
\hline
\end{tabular}
}

\end{center}
\end{table}

On the first step we learned the neural networks to reconstruct the signal
having, as well as training patterns, one Gaussian peak. Examples of initially
unknown reconstructed distributions with different variances and $n=5$ are
shown in Tables \ref{example-delta}, \ref{example-smooth}. The neural networks
was learned independently in all the cases with the similar randomly generated
patterns. 

\begin{table}
\caption{\label{example-delta2} The initial (a) and reconstructed by the neural
network (b) 
$\delta$-like distribution of physical quantity with two peaks.}

\begin{center}

\subtable[]{
\begin{tabular}{rrrrr}
\hline
  0&  0&  0&  0&  0\\                                   
  0&  0&  0.18609&  0&  0\\                                   
  0&  0&  0&  0&  0\\                                   
  0&  0&  0.12760&  0&  0\\
  0&  0&  0&  0&  0\\                           
\multicolumn{1}{c}{\rule{3.5em}{0ex}} & \multicolumn{1}{c}{\rule{3.5em}{0ex}} & 
\multicolumn{1}{c}{\rule{3.5em}{0ex}} & \multicolumn{1}{c}{\rule{3.5em}{0ex}} & 
\multicolumn{1}{c}{\rule{3.5em}{0ex}} \\[-\arraystretch pc]
\hline
\end{tabular}
}

\subtable[]{
\begin{tabular}{rrrrr}
\hline
  0.00125&  0.00011&  0.01166& -0.00362& -0.00194\\                                   
 -0.00072&  0.00442&  0.19818& -0.00477& -0.00122\\                                  
 -0.00901& -0.01231&  0.02648&  0.00080&  0.00025\\                                   
  0.00540&  0.00196&  0.09614&  0.00528& -0.00353\\                                   
 -0.00271&  0.00754&  0.00120&  0.00300& -0.00040\\                                   
\multicolumn{1}{c}{\rule{3.5em}{0ex}} & \multicolumn{1}{c}{\rule{3.5em}{0ex}} & 
\multicolumn{1}{c}{\rule{3.5em}{0ex}} & \multicolumn{1}{c}{\rule{3.5em}{0ex}} & 
\multicolumn{1}{c}{\rule{3.5em}{0ex}} \\[-\arraystretch pc]
\hline
\end{tabular}
}

\end{center}
\end{table}

\begin{table}
\caption{\label{example-smooth2} The initial (a) and reconstructed by the neural
network (b) 
smooth distribution of physical quantity with two peaks.}

\begin{center}

\subtable[]{
\begin{tabular}{rrrrr}
\hline
  0.00537&  0.02746&  0.07162&  0.09016&  0.04897\\                                   
  0.01875&  0.08318&  0.17714&  0.18591&  0.09016\\                                   
  0.02802&  0.11444&  0.20783&  0.17714&  0.07162\\                                   
  0.01754&  0.06892&  0.11444&  0.08318&  0.02746\\                                  
  0.00453&  0.01754&  0.02802&  0.01875&  0.00537\\                                   
\multicolumn{1}{c}{\rule{3.5em}{0ex}} & \multicolumn{1}{c}{\rule{3.5em}{0ex}} & 
\multicolumn{1}{c}{\rule{3.5em}{0ex}} & \multicolumn{1}{c}{\rule{3.5em}{0ex}} & 
\multicolumn{1}{c}{\rule{3.5em}{0ex}} \\[-\arraystretch pc]
\hline
\end{tabular}
}

\subtable[]{
\begin{tabular}{rrrrr}
\hline
  0.00781&  0.02640&  0.07310&  0.09046&  0.05154\\
  0.01913&  0.08428&  0.17919&  0.18420&  0.09284\\
  0.03088&  0.11705&  0.20816&  0.18198&  0.07528\\
  0.02103&  0.07112&  0.11420&  0.08583&  0.02474\\
  0.00516&  0.01640&  0.02912&  0.02363&  0.00758\\
\multicolumn{1}{c}{\rule{3.5em}{0ex}} & \multicolumn{1}{c}{\rule{3.5em}{0ex}} & 
\multicolumn{1}{c}{\rule{3.5em}{0ex}} & \multicolumn{1}{c}{\rule{3.5em}{0ex}} & 
\multicolumn{1}{c}{\rule{3.5em}{0ex}} \\[-\arraystretch pc]
\hline
\end{tabular}
}

\end{center}
\end{table}

On the next step we trained neural networks reconstructing distributions with
two Gaussian peaks. Examples are shown in the Tables \ref{example-delta2},
\ref{example-smooth2}. In the case of the smooth distribution the reconstructed
signal is close to the original. For the $\delta$-like distributions the
locations of the peaks are recovered correctly but their magnitude differs from
the original.

One can see from Tables that neural
network sufficiently accurately reconstructs unknown pattern and can be used in
practice.

Presented neural network unlike Refs. \cite{psr99,lb95} can work not only with
signal from fiber-optic measuring line being within the interval of linear
dependence of light intensity on phase shift.

\section{Conclusions}

The computer model of the neural network solving tomographic problem is
proposed. This network can be used in conjunction with the distributed
fiber-optic measuring system and partly may be made as electronic or optical
hardware.

\end{document}